\documentclass[12pt]{iopart}
\usepackage{epsfig}
\def\bea{\begin{equation}}
\def\eea{\end{equation}}
\def\bea{\begin{eqnarray}}
\def\eea{\end{eqnarray}}
\def\ba{\begin{array}}
\def\ea{\end{array}}
\def\slash#1{\setbox0=\hbox{$#1$}#1\hskip-\wd0\dimen0=5pt\advance
        \dimen0 by-\ht0\advance\dimen0 by\dp0\lower0.5\dimen0\hbox
          to\wd0{\hss\sl/\/\hss}}

\begin{document}

\begin{flushright}
WITS-MITP-013
\end{flushright}

\title[Hyperbolic extra-dimensions]{Hyperbolic extra-dimensions in particle physics and beyond}

\author{Alan S. Cornell}

\address{National Institute for Theoretical Physics, School of Physics and Mandelstam Institute for Theoretical Physics,
University of the Witwatersrand, Johannesburg, Wits 2050, South Africa}
\ead{alan.cornell@wits.ac.za}
\vspace{10pt}
\begin{indented}
\item[]July 2015
\end{indented}

\begin{abstract}
The nature of space-time at high energy is an open question and the link between extra-dimensional theories with the physics of the Standard Model can not be established in a unique way. The compactification path is not unique and supersymmetry breaking can be done in different ways. Compactifications based on hyperbolic orbifolds gather a large number of properties that are useful for this problem, like a Dirac spectrum chiral zero modes, a mass gap with the Kaluza-Klein modes, discrete residual symmetries for the stability of dark matter, and interesting cosmological constructions. This shall be explored here for bulk scalar fields, with a roadmap provided for fermionic studies and beyond.
\end{abstract}

\section{Introduction}

\par At the beginning of the twentieth century, the idea that spacetime could have more than four dimensions was put forward by Theodore Kaluza \cite{Kaluza:1921tu} and Oskar Klein \cite{Klein:1926tv}. What makes this idea of extra-dimensions so enduring is that they have not been ruled out by experiments, where we presume they are compact enough to have avoided detection thus far. In some scenarios they could even be large and still have avoid detection.

\par The idea of a large extra-dimensional model was first proposed by Arkani-Hamed, Dvali and Dimopoulos (ADD) \cite{ArkaniHamed:1998rs, Antoniadis:1998ig}, where their ADD model introduced the idea that the Standard Model (SM) fields are confined to a brane (a four-dimensional manifold) which resides in the full spacetime. Due to their confinement to the brane, the SM fields do not feel the effects of the extra-dimensions, and experimental results are independent of their size. However, gravity is not so constrained given that it has no SM charges, and can feel the entire space. As such, large deviations from Newton's inverse square law are possible at short distances, though due to gravity being weaker than the other forces, this has only been experimentally verified to distances of the order of micrometres. Therefore the limits placed on this gravitational law are very weak.

\par A primary motivation for studying extra-dimensional models was to resolve the hierarchy problem between the (effective) Planck scale, $M_{Pl} = {\cal O}(10^{18}GeV)$, and the electroweak scale, $M_{EW} = {\cal O}(100GeV)$. This stems from the expectation that the bare Higgs mass would obtain corrections of the order of $M_{Pl}$, meaning some extreme fine-tuning of the Higgs sector parameters is required for the electroweak scale to be so low. The ADD model can solve this by virtue of the true Planck scale being related to the four dimensional Planck scale via the volume of the extra-dimensions. This is the result of gravity being able to propagate in the full spacetime, such that the Planck scale we measure is effective and valid for energies smaller than the inverse radius of compactification of these extra-dimensions. As such, if the volume of the extra-dimensions is large enough, then the true scale of gravity can be as low as the electroweak scale. However, the ADD model's hierarchy problem resolution replaces the hierarchy between $M_{EW}$ and $M_{Pl}$ with another hierarchy, that between $M_{EW}$ and the inverse radius of compactification for the extra-dimensions. Thus this ADD resolution only raises another question, why is the inverse radius of compactification so large when compared to the electroweak scale?

\par The construction of the ADD model assumes that the extra-dimensions are flat, with compactification being on a torus. But what if we change this assumption? Could other geometries for this space avoid this issue? Kaloper {\it et al.} \cite{Kaloper:2000jb}, with this in mind, argued that a compact hyperbolic space may resolve this concern. At this point I note that other extra-dimensional models which contain branes do exist, for example the Randall-Sundrum model \cite{Randall:1999vf, Randall:1999ee} where two branes were used in its construction. We will not consider these models here.

\par In these proceedings we shall consider, as a preliminary step, the dynamics of a bulk scalar field in a hyperbolic extra-dimensional model and how to determine its Kaluza-Klein (KK) spectrum. We shall then discuss possible extensions of this to fermion fields, and ultimately constructing a realistic model with all SM like fields and interactions.

\section{Scalar Eigenmodes}

\par As was discussed in the introduction, the ADD model resolved the hierarchy problem between $M_{Pl}$ and $M_{EW}$ by replacing it with a question as to why the radius of the extra-dimensional space is so large compared to the electroweak scale. If, however, the extra-dimensions were a hyperbolic disk with radius $R$, its volume would be:
\begin{equation}
V = \frac{4\pi}{\lambda^2} \sinh^2 \left( \frac{\lambda R}{2} \right) \sim \frac{\pi}{\lambda^2} e^{\lambda R} \; ,
\end{equation}
where $\lambda$ is the constant negative curvature of our hyperbolic disk. This has the result that it is now possible to have $M_{Pl} \sim M_{EW} \sim R^{-1}$, that is, no new hierarchy is introduced if $\lambda$ is of the order of the fundamental scale of gravity \cite{Melbeus:2008hk}.

\par As there exist a range of possible hyperbolic geometries, we need to study which can provide us with the most interesting phenomenological features (to possibly provide dark matter candidates etc.). To begin this probe of possible spaces we look at scalar fields, where a great deal of information about the geometry of hyperbolic spaces can be obtained from the eigenmodes of the Laplace operator \cite{Cornish}. For hyperbolic spaces we have the Mostow-Prasad rigidity theorem \cite{Mostow} which ensures that different spaces have different eigenvalue spectra.

\par Note, though, that the eigenmodes of hyperbolic spaces cannot be expressed in a closed analytic form, as such numerical solutions are used. Of the numerical methods which exist the boundary element method of Aurich and Steiner \cite{Aurich} is the most powerful, however, here we shall use an alternative approach of Cornish and Spergel \cite{Cornish}. We appreciate that this method is technically weaker, but it is more adaptable to studying a variety of hyperbolic spaces as the only inputs are the group generators.

\par This approach is based on a grid over the extra-dimensional space, where we need a coordinate system on the hyperbolic space that allows us to make a discretisation in a sensible way. For example, a possible metric on a hyperbolic space, written in spherical coordinates, can be:
\begin{equation}
ds^2 = d\rho^2 + \sinh^2\rho \left( d\theta^2 + \sin^2 \theta d\phi^2 \right) \; . \label{metric}
\end{equation}
In these coordinates the Laplace operator acting on a scalar field $\Phi$ is
\begin{eqnarray}
\Delta \Phi & = & \frac{1}{\sinh^2 \rho} \Bigg[ \frac{\partial}{\partial \rho} \left( \sinh^2 \rho \frac{\partial \Phi}{\partial \rho} \right) + \frac{1}{\sin^2 \theta} \frac{\partial}{\partial \theta} \left( \sin \theta \frac{\partial \Phi}{\partial \theta} \right) + \frac{1}{\sin^2 \theta} \frac{\partial^2 \Phi}{\partial \phi^2} \Bigg] \; .
\end{eqnarray}
In the simply connected space the eigenvalues take all values in the range $q^2 = [1,\infty)$, and the eigenmodes are given by
\begin{equation}
Q_{q\ell m} (\rho, \theta, \phi) = X_q^\ell(\rho) Y_{\ell m}(\theta, \phi) \; .
\end{equation}
The $Y_{\ell m}$'s are spherical harmonics and the radial eigenfunctions are given by hyperspherical Bessel functions. The wavenumber, $k$, is related to the eigenvalues of the Laplacian by $k^2 = q^2 - 1$, and the modes have wavelength $2\pi/k$. These eignemodes are also normalised.

\par As noted in Ref. \cite{Cornish}, the eigenmodes can be lifted to the universal cover and expressed in terms of the eigenmodes of $H^3$:
\begin{equation}
\Phi_q = \sum^\infty_{\ell = 0} \sum^\ell_{m = - \ell} a_{q\ell m} Q_{q \ell m} \; .
\end{equation}
The modes $\Phi_q$ must satisfy the property
\begin{equation}
\Phi_q (x) = \Phi_q (g x) \; \forall g \in \Gamma \; \mathrm{and} \; \forall x \in H^3 \; ,
\end{equation}
which places restrictions on the expansion coefficients $a_{q\ell m}$. As such, this equation can only be satisfied when $q^2$ is an eigenvalue of the compact space. To find these eigenmodes we solve numerically.
\begin{figure}[h]
\centering
\includegraphics[width=8cm]{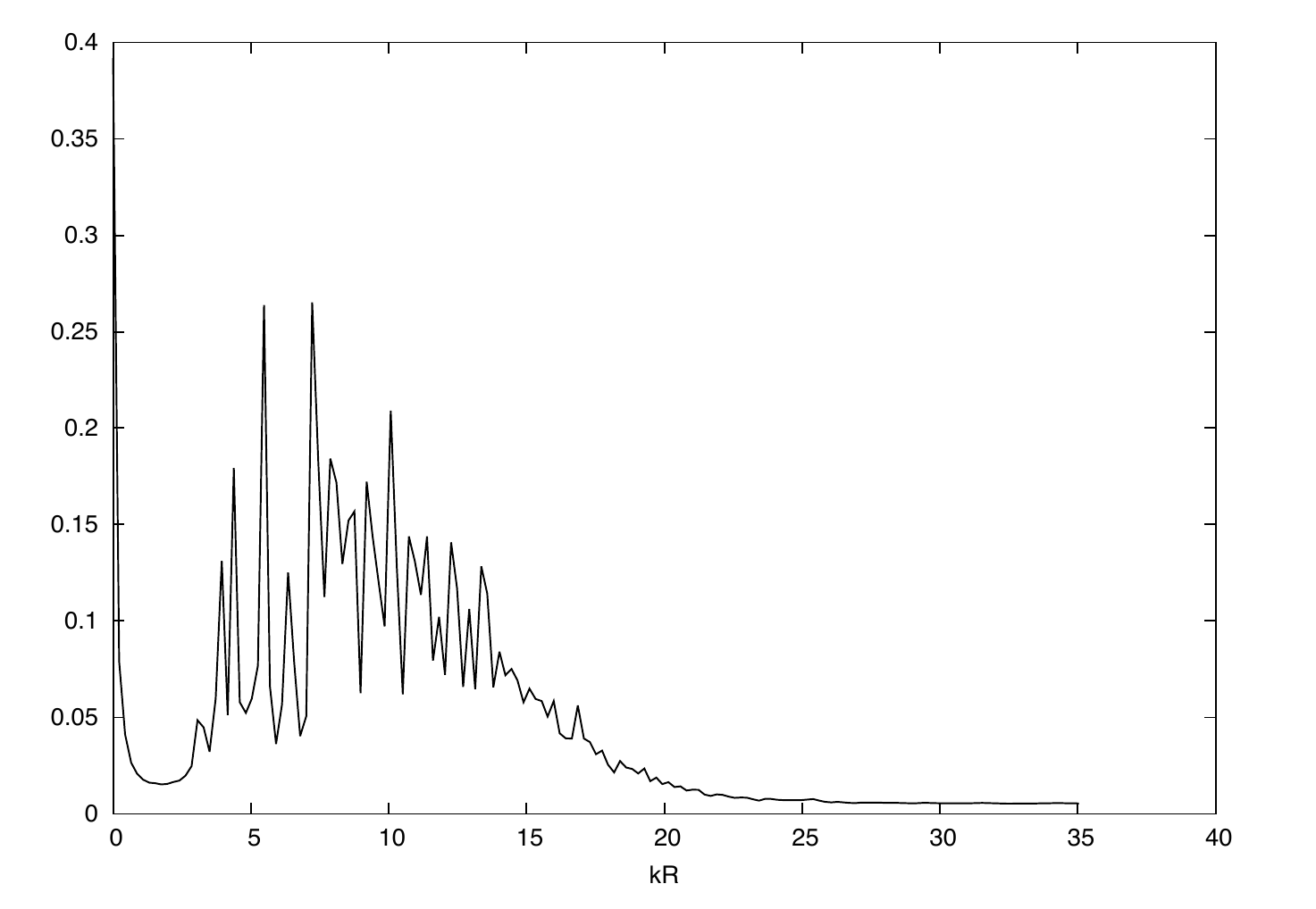}
\caption{\it The KK spectrum for $m188(-1,1)$ in the range $kR \to 35$.}\label{fig:1}
\end{figure}
\par As a test case we follow the example of Ref. \cite{Cornish} and choose from the {\it SnapPea} \cite{Weeks} census of closed hyperbolic 3-manifolds for the face-pairing generators $m188(-1, 1)$. We ran our code out to $kR = 35$ to the produce the plot in figure \ref{fig:1}. This figures shows an interesting feature common to these spaces, that of a zero mode followed by a mass gap, and then many modes after that which are nearly degenerate. Phenomenologically this could provide a wealth of dark matter states, which we are currently investigating further.

\begin{figure}[h]
\centering
\includegraphics[width=7cm]{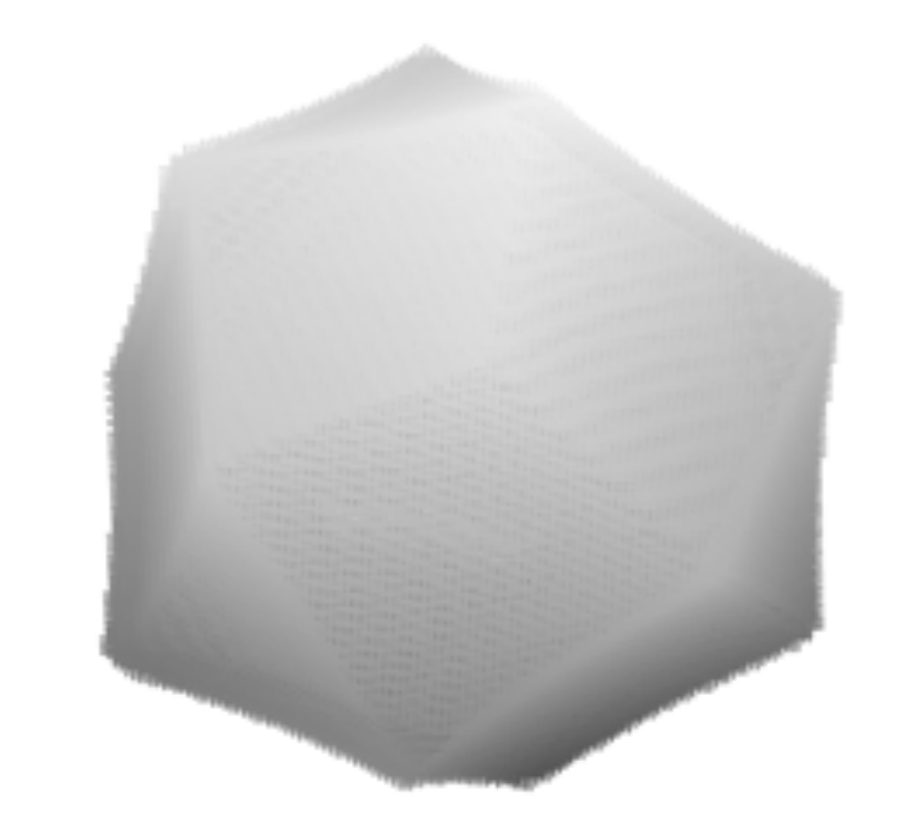}
\caption{\it A Dirichlet domain for the space chosen above shown in Klein coordinates.}\label{fig:2}
\end{figure}
\par For those curious to see what the eigenmodes look like, we display in figure \ref{fig:2} a view of the Dirichlet domain, to help make contact with the 3-dimensional structure of the modes.

\section{Future extensions}

\par In order to now extend this to fermions in a bulk hyperbolic space, we first write the derivative part of the Dirac operator in Cartesian coordinates:
\begin{equation}
\slash\partial = \sigma^a e^\mu_a \partial_\mu = \frac{1}{2} \left( 1 - \frac{r^2}{L^2} \right) \left( \begin{array}{cc}
\partial_z & \partial_x - i \partial_y \\ 
\partial_x + i \partial_y & - \partial_z 
\end{array} \right) \; .
\end{equation}
We have here employed a nice compact (though not manifestly covariant) form for the dreibein
\begin{equation}
e^\mu_a = \frac{1}{2} \left( 1 - \frac{r^2}{L^2} \right) \delta^\mu_a \; , 
\end{equation}
using our earlier example metric, equation (\ref{metric}), which gives the tangent frame
\begin{equation}
\Theta^a_\mu = \frac{2}{1 - r^2/L^2} \delta^a_\mu 
\end{equation}
and spin connection
\begin{equation}
\omega^{ij}_\mu = - x^{[i}\Theta^{j]}_\mu /L^2 \; .
\end{equation}
These shall be necessary for the construction of the appropriate Dirac equations, which we look forward to reporting on further soon.

\par Note, however, that a KK spectrum, as was observed in the scalar case, would be extremely interesting, phenomenologically, due to the possibility of chiral zero modes, as we know to be possible in six-dimensional theories \cite{Cacciapaglia:2011hx}. Such chiral modes are not possible in five-dimensional theories, and many other extra-dimensional scenarios, but are crucial to the construction of the SM.

\par To conclude, the nature of space-time at high energy is an open question and the link between extra-dimensional theories with the physics of the SM can not be established in a unique way. The compactification path is not unique and in this proceeding we have sought to tackle this problem by starting from what is known from theory and experiment: the SM contains chiral fermions, the dark matter content of the universe and the difference between the electroweak and Planck scale should be explained. Compactifications based on hyperbolic orbifolds gather a large number of properties that are useful for these problems, like a Dirac spectrum chiral zero modes, a mass gap with the KK modes, discrete residual symmetries for the stability of dark matter, and interesting cosmological constructions \cite{Kim:2010fq, Orlando:2010kx}. However, a complete model involving all SM-like fields is still to be constructed, where we hope to report on more progress along these lines at the next SAIP meeting.

\section*{Acknowledgements}

\par I would like to thank my collaborators at IPNL, Aldo Deandrea, Giacomo Cacciapaglia and Nicolas Deustchmann, for their hospitality and useful discussions. These proceedings are based on preliminary results from a paper we are in the process of writing.


\end{document}